\documentclass[twocolumn,pre,superscriptaddress,lengthcheck]{revtex4}

\usepackage{dcolumn}
\usepackage{amsmath}
\usepackage[bf]{subfigure}

\usepackage{graphicx}
\usepackage{rotating}

\newcommand{\av}[1]{\langle#1\rangle}
\newcommand{\etal}{{\it{}et~al.}}
\newcommand{\defn}{\textit}
\newcommand{\Ord}{\mathrm{O}}
\newcommand{\Tr}{\mathop\mathrm{Tr}}

\newcommand{\mat}{\mathbf}
\renewcommand{\vec}{\mathbf}

\begin{document}

\title{Localization and centrality in networks}
\author{Travis Martin}
\email[Please direct correspondence to: ]{travisbm@umich.edu}
\affiliation{Department of Electrical Engineering and Computer Science,
  University of Michigan, Ann Arbor, MI 48109}
\author{Xiao Zhang}
\affiliation{Department of Physics, University of Michigan, Ann Arbor, MI
  48109}
\author{M. E. J. Newman}
\affiliation{Department of Physics, University of Michigan, Ann Arbor, MI
48109}
\affiliation{Center for the Study of Complex Systems, University of
Michigan, Ann Arbor, MI 48109}

\begin{abstract}
  Eigenvector centrality is a common measure of the importance of nodes in
  a network.  Here we show that under common conditions the eigenvector
  centrality displays a localization transition that causes most of the
  weight of the centrality to concentrate on a small number of nodes in the
  network.  In this regime the measure is no longer useful for
  distinguishing among the remaining nodes and its efficacy as a network
  metric is impaired.  As a remedy, we propose an alternative centrality
  measure based on the nonbacktracking matrix, which gives results closely
  similar to the standard eigenvector centrality in dense networks where
  the latter is well behaved, but avoids localization and gives useful
  results in regimes where the standard centrality fails.
\end{abstract}

\maketitle

\section{Introduction}
In the study of networked systems such as social, biological, and
technological networks, centrality is one of the most fundamental of
metrics.  Centrality quantifies how important or influential a node is
within a network.  The simplest of centrality measures, the \emph{degree
  centrality}, or simply degree, is the number of connections a node has to
other nodes.  In a social network of acquaintances, for example, someone
who knows many people is likely to be more influential than someone who
knows few or none.  \defn{Eigenvector centrality}~\cite{Bonacich72a} is a
more sophisticated variant of the same idea, which recognizes that not all
acquaintances are equal.  You are more influential if the people you know
are themselves influential.  Eigenvector centrality defines a centrality
score~$v_i$ for each node~$i$ in an undirected network, which is
proportional to the sum of the scores of the node's network neighbors $v_i
= \lambda^{-1} \sum_j A_{ij} v_j$, where $\lambda$ is a constant and the
sum is over all nodes.  Here $A_{ij}$ is an element of the adjacency
matrix~$\mat{A}$ of the network having value one if there is an edge
between nodes $i$ and~$j$ and zero otherwise.  Defining a vector~$\vec{v}$
whose elements are the $v_i$, we then have $\mat{A}\vec{v} =
\lambda\vec{v}$, meaning that the vector of centralities is an eigenvector
of the adjacency matrix.  If we further stipulate that the centralities
should all be nonnegative, it follows by the Perron--Frobenius
theorem~\cite{Strang09} that $\vec{v}$ must be the leading eigenvector (the
vector corresponding to the most positive eigenvalue~$\lambda$).
Eigenvector centrality and its variants are some of the most widely used of
all centrality measures.  They are commonly used in social network
analysis~\cite{WF94} and form the basis for ranking algorithms such as the
HITS algorithm~\cite{Kleinberg99a} and the eigenfactor metric~\cite{BWW08}.

As we argue in this paper, however, eigenvector centrality also has serious
flaws.  In particular, we show that, depending on the details of the
network structure, the leading eigenvector of the adjacency matrix can
undergo a localization transition in which most of the weight of the vector
concentrates around one or a few nodes in the network.  While there may be
situations, such as the solution of certain physical models on networks, in
which localization of this kind is useful or at least has some scientific
interest, in the present case it is undesirable, significantly diminishing
the effectiveness of the centrality as a tool for quantifying the
importance of nodes.  Moreover, as we will show, localization can happen
under common real-world conditions, for instance in networks with power-law
degree distributions.

As a solution to these problems, we propose a new centrality measure based
on the leading eigenvector of the Hashimoto or nonbacktracking
matrix~\cite{Hashimoto89,Krzakala13}. This measure has the desirable
properties of (1)~being closely equal to the standard eigenvector
centrality in dense networks, where the latter is well behaved, while also
(2)~avoiding localization, and hence giving useful results, in cases where
the standard centrality fails.

\section{Localization of eigenvector centrality}
A number of numerical studies of real-world networks have shown evidence of
localization phenomena in the past~\cite{FDBV01,GKK01a,ESMS03,CM11,GDOM12}.
In this paper we formally demonstrate the existence of a localization phase
transition in the eigenvector centrality and calculate its properties using
techniques of random matrix theory.

The fundamental cause of the localization phenomenon we study is the
presence of ``hubs'' within networks, nodes of unusually high degree, which
are a common occurrence in many real-world networks~\cite{BA99b}.  Consider
the following simple undirected network model consisting of a random graph
plus a single hub node, which is a special case of a model introduced
previously in~\cite{NN13}.  In a network of $n$ nodes, $n-1$~of them form
a random graph in which every distinct pair of nodes is connected by an
undirected edge with independent probability~$c/(n-2)$, where $c$ is the
mean degree.  The $n$th node is the hub and is connected to every other
node with independent probability $d/(n-1)$, so that the expected degree of
the hub is~$d$.  In the regime where $c\gg1$ it is known that (with high
probability) the spectrum of the random graph alone has the classic Wigner
semicircle form, centered around zero, plus a single leading eigenvalue
with value $c+1$ and corresponding leading eigenvector equal to the uniform
vector $(1,1,1,\ldots)/\sqrt{n}$ plus random Gaussian noise of
width~$\Ord(1/\sqrt{n})$~\cite{Tao12}. Thus the eigenvector centralities of
all vertices are $\Ord(1/\sqrt{n})$ with only modest fluctuations.  No
single node dominates the picture and the eigenvector centrality is well
behaved.

If we add the hub to the picture, however, things change.  The addition of
an extra vertex naturally adds one more eigenvalue and eigenvector to the
spectrum, whose values we can calculate as follows.  Let $\mat{X}$ denote
the $(n-1)\times(n-1)$ adjacency matrix of the random graph alone and let
the vector~$\vec{a}$ be the first $n-1$ elements of the final row and
column, representing the hub.  (The last element is zero.)  Thus the
full adjacency matrix has the form
\begin{equation}
\mat{A} = \begin{pmatrix}
  \boxed{\begin{matrix} \\ \\ \qquad\,\mat{X}\,\qquad{} \\ \\ \\ \end{matrix}}
  &
  \boxed{\begin{matrix} \\ \phantom{m} \\ \vec{a} \\ \\ \null \end{matrix}}\,{}
  \\
  \boxed{\qquad\vec{a}^T\qquad} & 0\rule{2pt}{0pt}\rule{0pt}{16pt} \\
          \end{pmatrix}.
\end{equation}

Let $z$ be an eigenvalue of~$\mat{A}$ and let
$\vec{v}=(\vec{v}_1|v_n)$ be the corresponding eigenvector, where
$\vec{v}_1$ represents the first $n-1$ elements and $v_n$ is the last
element.  Then, multiplying out the eigenvector equation $\mat{A}\vec{v} =
z\vec{v}$, we find
\begin{equation}
\mat{X}\vec{v}_1 + v_n\vec{a} = z\vec{v}_1, \qquad
\vec{a}^T\vec{v}_1 = zv_n.
\end{equation}
Rearranging the first of these, we get
\begin{equation}
\vec{v}_1 = v_n(z\mat{I}-\mat{X})^{-1}\vec{a},
\label{eq:v1}
\end{equation}
and substituting into the second we get
\begin{equation}
\vec{a}^T (z\mat{I}-\mat{X})^{-1} \vec{a} = z,
\label{eq:hub}
\end{equation}
where $\mat{I}$ is the identity.  Writing the matrix inverse in terms of
its eigendecomposition $(z\mat{I}-\mat{X})^{-1} = \sum_i
\vec{x}_i(z-\chi_i)^{-1}\vec{x}_i^T$, where $\vec{x}_i$ is the $i$th
eigenvector of~$\mat{X}$ and $\chi_i$ is the corresponding eigenvalue,
Eq.~\eqref{eq:hub} becomes
\begin{equation}
{(\vec{a}^T\vec{x}_1)^2\over z-(c+1)} 
  + \sum_{i=2}^{n-1} {(\vec{a}^T\vec{x}_i)^2\over z-\chi_i} = z,
\label{eq:hub2}
\end{equation}
where we have explicitly separated the largest eigenvalue~$\chi_1=c+1$ and
the remaining $n-2$ eigenvalues, which follow the semicircle law.

\begin{figure}
\begin{center}
\includegraphics[width=8.5cm]{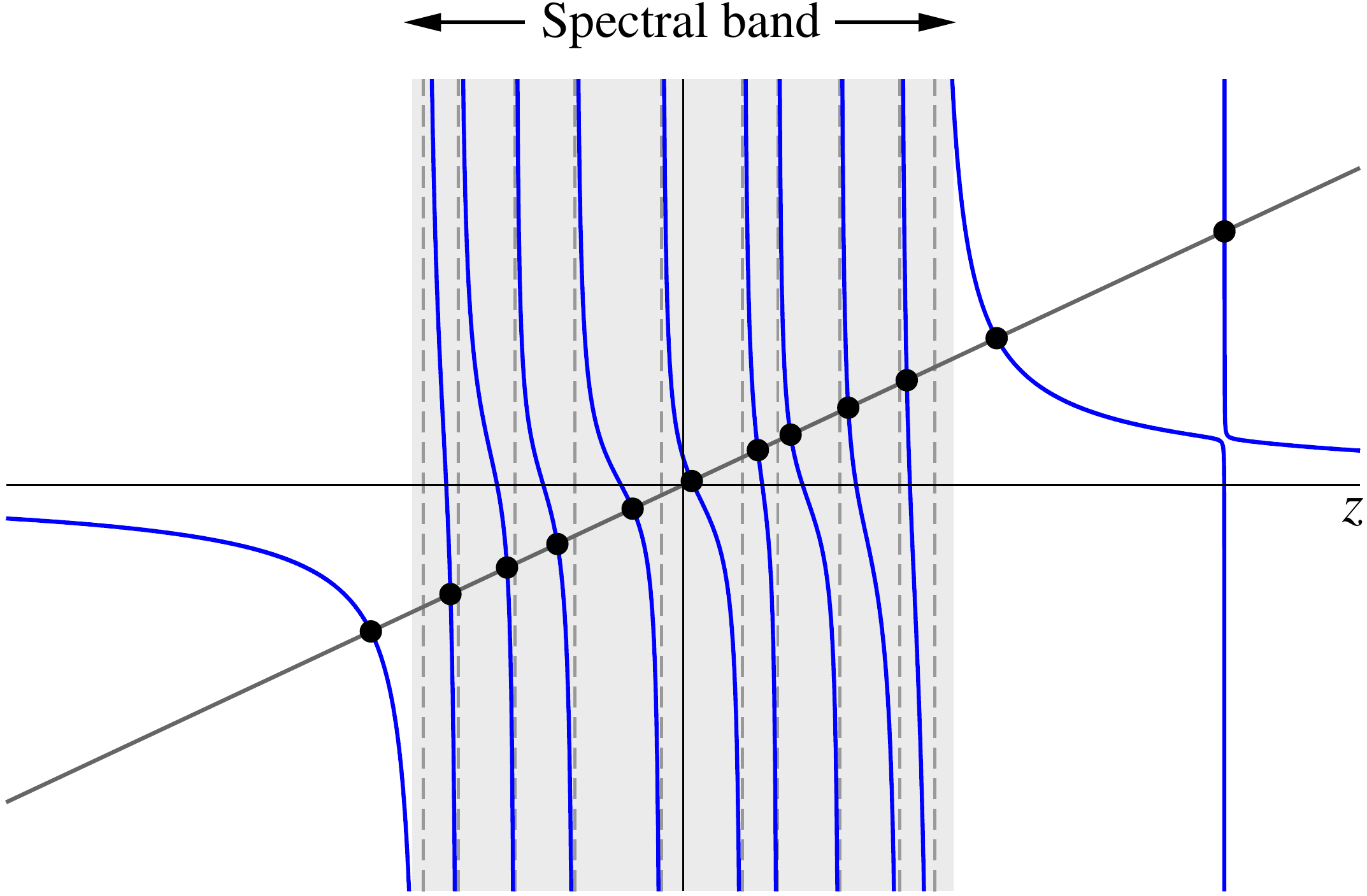}
\end{center}
\caption{(Color online) Graphical representation of the solution of
  Eq.~\eqref{eq:hub2}.  The curves represent the left-hand side of the
  equation, which has poles at the positions of the eigenvalues~$\chi_i$
  (marked by the vertical dashed lines).  The diagonal line represents the
  right-hand side and the points where the two cross, marked by dots, are
  the solutions of the equation for~$z$.}
\label{fig:solution}
\end{figure}

Although we don't know the values of the quantities $\vec{a}^T\vec{x}_i$
appearing in Eq.~\eqref{eq:hub2}, the left-hand side as a function of~$z$
clearly has poles at each of the eigenvalues~$\chi_i$ and a tail that goes
as $1/z$ for large~$z$.  Moreover, for properly normalized~$\vec{x}_1$ the
numerator of the first term in the equation is $\Ord(1/n)$ and hence this
term diverges significantly only when $z-(c+1)$ is also~$\Ord(1/n)$,
i.e.,~when $z$ is very close to the leading eigenvalue~$c+1$.  Hence the
qualitative form of the function must be as depicted in
Fig.~\ref{fig:solution} and solutions to the full equation correspond to
the points where this form crosses the diagonal line representing the
right-hand side of the equation.  These points are marked with dots in the
figure.

As the geometry of the figure makes clear, the solutions for~$z$, which are
the eigenvalues of the full adjacency matrix of our model including the hub
vertex, must fall in between the eigenvalues~$\chi_i$ of the
matrix~$\mat{X}$, and hence satisfy an interlacing condition of the form
$z_1>\chi_1>z_2>\chi_2>\ldots>\chi_{n-1}>z_n$, where we have numbered both
sets of eigenvalues in order from largest to smallest.  In the limit where
the network becomes large and the eigenvalues $\chi_2\ldots\chi_{n-1}$ form
a continuous semicircular band, this interlacing imposes tight bounds on
the solutions $z_3$ to~$z_{n-1}$, such that they must follow the same
semicircle distribution.  Moreover, the leading eigenvalue~$z_1$ has to fall
within $\Ord(1/n)$ of $\chi_1=c+1$, and hence $z_1\to c+1$ in the large
size limit.

This leaves just two unknown eigenvalues, $z_2$~lying above the
semicircular band and $z_n$ lying below it.  In the context of the
eigenvector centrality it is the one at the top that we care about.  In
Fig.~\ref{fig:solution} this eigenvalue is depicted as lying below the
leading eigenvalue~$z_1$, but it turns out that this is not always the
case, as we now show.

Consider Eq.~\eqref{eq:hub2} for any value of~$z$ well away from~$c+1$, so
that the first term on the left can be neglected (meaning that $z$ is not
within $\Ord(1/n)$ of $c+1$).  The vector~$\vec{x}_i$ for $i\ge2$ is
uncorrelated with $\vec{a}$ and hence the product $\vec{a}^T\vec{x}_i$ is a
Gaussian random variable with variance~$d/n$ and, averaging over the
randomness, the equation then simplifies to
\begin{equation}
{d\over n} \Tr(z\mat{I}-\mat{X})^{-1} = z.
\label{eq:trace}
\end{equation}
The quantity $g(z) = n^{-1} \Tr(z\mat{I}-\mat{X})^{-1}$ is a standard one
in the theory of random matrices---it is the so-called Stieltjes transform
of~$\mat{X}$, whose value for a symmetric matrix with iid elements such as
this one is known to be~\cite{Tao12}
\begin{equation}
g(z) = {z - \sqrt{z^2-4c}\over2c}.
\label{eq:stieltjes}
\end{equation}
Combining Eqs.~\eqref{eq:trace} and~\eqref{eq:stieltjes} and solving for
$z$ we find the eigenvalue we are looking for:
\begin{equation}
z_2 = {d\over\sqrt{d-c}}.
\end{equation}

Depending on the degree~$d$ of the hub, this eigenvalue may be either
smaller or larger than the other high-lying eigenvalue~$z_1=c+1$.  Writing
$d/\sqrt{d-c} > c+1$ and rearranging, we see that the hub eigenvalue
becomes the leading eigenvalue when
\begin{equation}
d > c(c+1),
\label{eq:threshold}
\end{equation}
i.e.,~when the hub degree is roughly the square of the mean degree.  Below
this point, the leading eigenvalue is the same as that of the random graph
without the hub and the eigenvector centrality is given by the
corresponding eigenvector, which is well behaved, so the centrality has no
problems.  Above this point, however, the leading eigenvector is the one
introduced by the hub, and this eigenvector, as we now show, has severe
problems.

If the eigenvector~$\vec{v}=(\vec{v}_1|v_n)$ is normalized to unity then
Eq.~\eqref{eq:v1} implies that
\begin{equation}
1 = |\vec{v}_1|^2 + v_n^2
  = v_n^2 \bigl[ \vec{a}^T(z\mat{I}-\mat{X})^{-2}\vec{a}+1 \bigr],
\end{equation}
and hence
\begin{align*}
v_n^2 &= {1\over\vec{a}^T(z\mat{I}-\mat{X})^{-2}\vec{a}+1}
       = {1\over(d/n) \Tr(z\mat{I}-\mat{X})^{-2}+1} \nonumber\\
      &= {1\over -dg'(z)+1},
\end{align*}
where $g(z)$ is again the Stieltjes transform, Eq.~\eqref{eq:stieltjes},
and $g'(z)$ is its derivative.  Performing the derivative and setting $z =
d/\sqrt{d-c}$, we find that
\begin{equation}
  v_n^2 = {d-2c\over2d-2c},
\end{equation}
which is constant and does not vanish as $n\to\infty$.  In other words, a
finite fraction of the weight of the vector is concentrated on the hub
vertex.

The neighbors of the hub also receive significant weight: the average of
their values is given by
\begin{equation}
{\vec{a}^T\vec{v}_1\over d} = {v_n\over d}
  \vec{a}^T(z\mat{I}-\mat{X})^{-1}\vec{a} = v_n g(z) = {v_n\over\sqrt{d-c}}.
\end{equation}
Thus they are smaller than the hub centrality~$v_n$, but still constant for
large~$n$.  Finally, defining the $(n-1)$-element uniform vector
$\vec{1}=(1,1,1,\ldots)$, the average of all $n-1$ non-hub vector elements
is
\begin{equation}
\av{v_i} = {\vec{1}^T\vec{v}_1\over{n-1}} = {v_n\over{n-1}}
  \vec{1}^T(z\mat{I}-\mat{X})^{-1}\vec{a},
\end{equation}
where we have used Eq.~\eqref{eq:v1} again.  Averaging over the randomness
and noting that $\mat{X}$ and $\vec{a}$ are independent and that the
average of~$\vec{a}$ is $d\vec{1}/(n-1)$, we then get
\begin{equation}
\av{v_i} = {dv_n\over n-1} g(z) = {1\over n-1}\, {dv_n\over\sqrt{d-c}},
\label{eq:avvi}
\end{equation}
which falls off as $1/n$ for large~$n$.

Thus, in the regime above the transition defined by~\eqref{eq:threshold},
where the eigenvector generated by adding the hub is the leading
eigenvector, a non-vanishing fraction of the eigenvector centrality falls
on the hub vertex and its neighbors, while the average vertex in the
network gets only an $\Ord(1/n)$ vanishing fraction in the limit of
large~$n$, much less than the $\Ord(1/\sqrt{n})$ fraction received by the
average vertex below the transition.  This is the phenomenon we refer to as
localization: the abrupt focusing of essentially all of the centrality on
just a few vertices as the degree of the hub passes above the critical
value~$c(c+1)$.  In the localized regime the eigenvector centrality picks
out the hub and its neighbors clearly, but assigns vanishing weight to the
average node.  If our goal is to determine the relative importance of
non-hub nodes, the eigenvector centrality will fail in the localized
regime.

\subsection{Numerical results}
\begin{figure}
\begin{center}
\includegraphics[width=8.5cm]{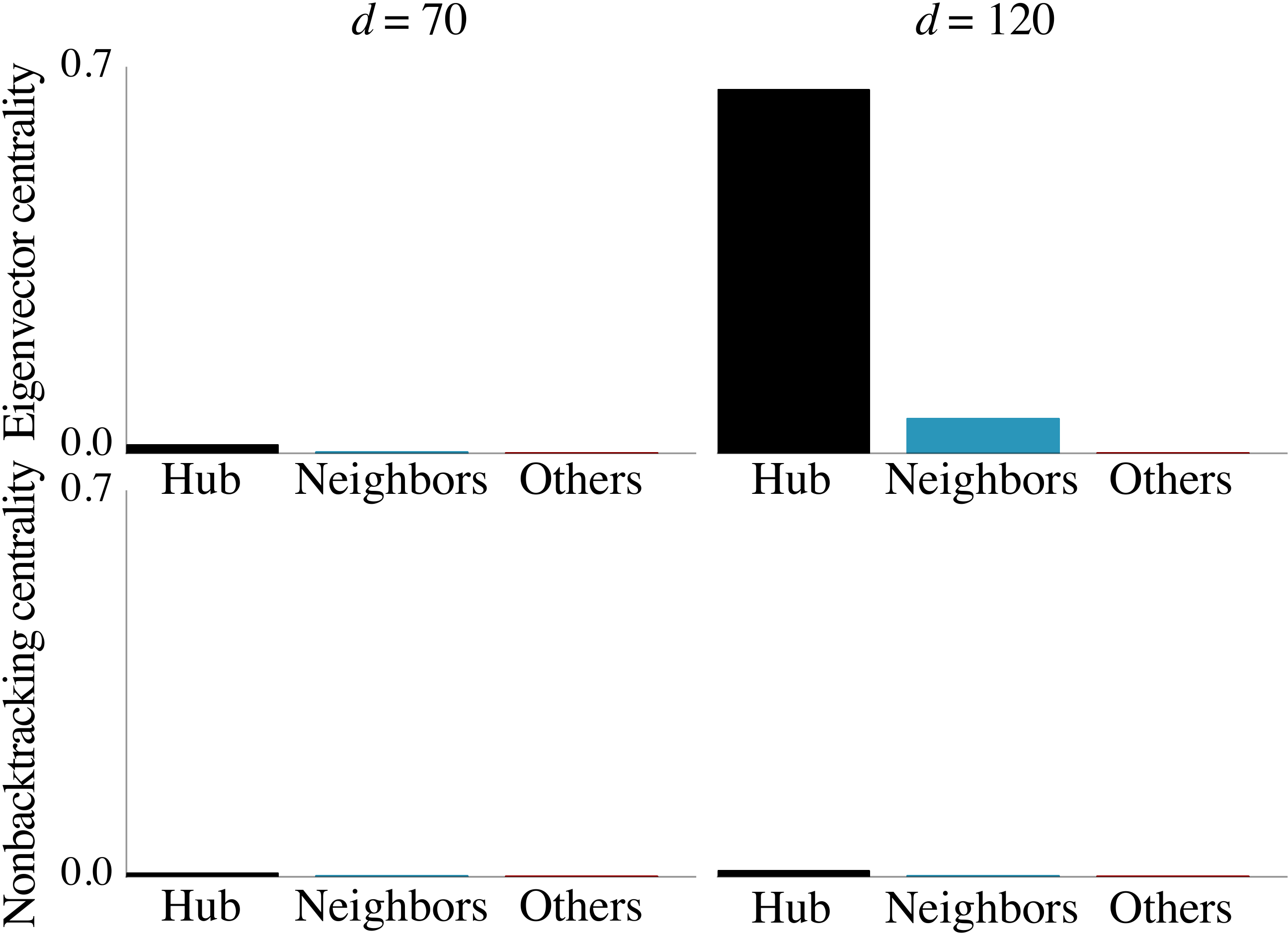}
\end{center}
\caption{(Color online) Bar charts of centralities for three categories of
  node for four examples of the model network studied here, as described in
  the text.  All plots share the same scale.  Error bars are small enough
  to be invisible on this scale.}
\label{fig:hub}
\end{figure}

As a demonstration of the localization phenomenon, we show in
Fig.~\ref{fig:hub} plots of the centralities of nodes in networks generated
using our model.  Each plot shows the average centrality of the hub, its
neighbors, and all other nodes for a one-million-node network with $c=10$.
The top two plots show the situation for the standard eigenvector
centrality for two different values of the hub degree---$d=70$ and $d=120$.
The former lies well within the regime where there is no localization,
while the latter is in the localized regime.  The difference between the
two is striking---in the first the hub and its neighbors get higher
centrality, as they should, but only modestly so, while in the second the
centrality of the hub vertex becomes so large as to dominate the figure.

\begin{figure}
\begin{center}
\includegraphics[width=8.5cm]{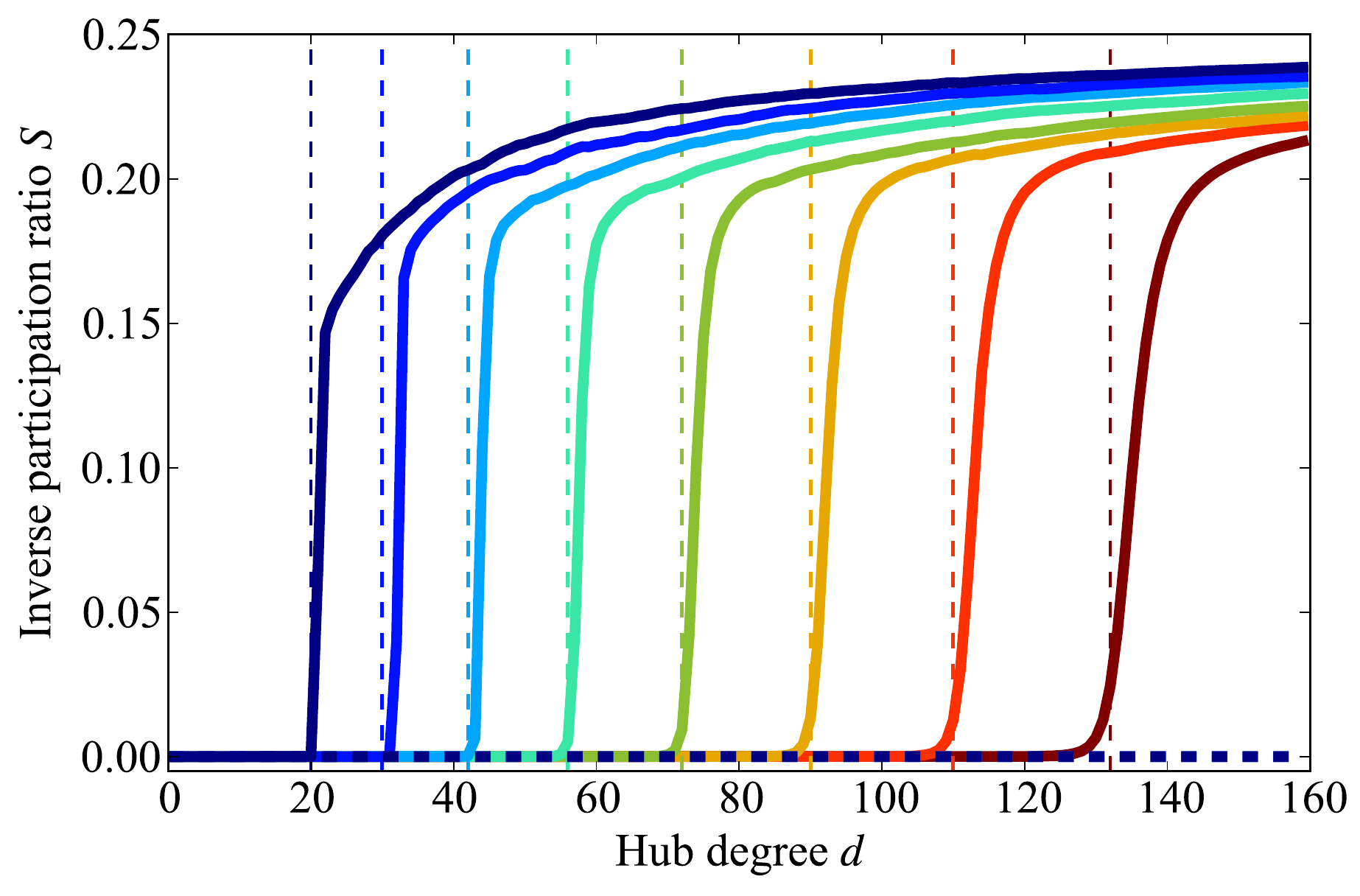}
\end{center}
\caption{(Color online) Numerical results for the inverse participation
  ratio~$S$ as a function of hub degree~$d$ for networks generated using
  the model described in the text with $n=1\,000\,000$ vertices and average
  degree $c$ ranging from 4 to~11.  The solid curves are eigenvector
  centrality; the horizontal dashed curves are the nonbacktracking
  centrality.  The vertical dashed lines are the expected positions of the
  localization transition for each curve, from Eq.~\eqref{eq:threshold}.}
\label{fig:transition}
\end{figure}

The extent of the localization can be quantified by calculating an inverse
participation ratio $S =\sum_{i=1}^n\,v_i^4$.  In the regime below the
transition where there is no localization and all elements~$v_i$ are
$\Ord(1/\sqrt{n})$ we have $S=\Ord(1/n)$.  But if one or more elements are
$\Ord(1)$, then $S=\Ord(1)$ also.  Hence if there is a localization
transition in the network then, in the limit of large~$n$, $S$~will go from
being zero to nonzero at the transition in the classic manner of an order
parameter.  Fig.~\ref{fig:transition} shows a set of such transitions in
our model, each falling precisely at the expected position of the
localization transition.

\subsection{Power-law networks}
So far we have looked only at the localization process in a simple model
network, but localization occurs in more realistic networks as well.  In
general, we expect it to be a problem in networks with high-degree hubs or
in very sparse networks, those with low average degree~$c$, where it is
relatively easy for the degree of a typical vertex to exceed the
localization threshold.  Many real-world networks fall into these
categories.  Consider, for example, the common case of a network with a
power-law degree distribution, such that the fraction~$p_k$ of nodes with
degree~$k$ goes as $k^{-\alpha}$ for some constant
exponent~$\alpha$~\cite{BA99b}. We can mimic such a network using the
so-called configuration model~\cite{MR95,NSW01}, a random graph with
specified degree distribution.  There are again two different ways a
leading eigenvalue can be generated, one due to the average behavior of the
entire network and one due to hub vertices of particularly high degree.  In
the first case the highest eigenvalue for the configuration model is known
to be equal to the ratio of the second and first moments of the degree
distribution~$\av{k^2}/\av{k}$ in the limit of large network size and large
average degree~\cite{CLV03,NN13}. At the same time, the leading eigenvalue
must satisfy the Rayleigh bound $z \ge
\vec{x}^T\mat{A}\vec{x}/\vec{x}^T\vec{x}$ for any real vector~$\vec{x}$,
with better bounds achieved when~$\vec{x}$ better approximates the true
leading eigenvector.  If $d$ denotes the highest degree of any hub in the
network and we choose an approximate eigenvector of form similar to the one
in our earlier model network, having elements $x_i=1$ for the hub,
$1/\sqrt{d}$~for neighbors of the hub, and zero otherwise, then the
Rayleigh bound implies $z\ge\sqrt{d}$.  Thus the eigenvector generated by
the hub will be the leading eigenvector whenever $\sqrt{d}>\av{k^2}/\av{k}$
(possibly sooner, but not later).

In a power-law network with $n$ vertices and exponent~$\alpha$, the highest
degree goes as $d\sim n^{1/(\alpha-1)}$~\cite{DMS01c} and hence increases
with increasing~$n$, while $\av{k^2}\sim d^{3-\alpha}$ and
$\av{k}\sim\textrm{constant}$ for the common case of $\alpha<3$.  Thus we
will have $\sqrt{d}>\av{k^2}/\av{k}$ for large~$n$ provided
$\frac12>3-\alpha$.  So we expect the hub eigenvector to dominate and the
eigenvector centrality to fail due to localization when $\alpha>\frac52$,
something that happens in many real-world networks.  (Similar arguments
have also been made by Chung~\etal~\cite{CLV03} and by
Goltsev~\etal~\cite{GDOM12}.)  We give empirical measurements of
localization in a number of real-world networks in Table~\ref{tab:power}
below.

\section{Nonbacktracking centrality}
So if eigenvector centrality fails to do its job, what can we do to fix it?
Qualitatively, the localization effect arises because a hub with high
eigenvector centrality gives high centrality to its neighbors, which in
turn reflect it back again and inflate the hub's centrality.  We can make
the centrality well behaved again by preventing this reflection.  To
achieve this we propose a modified eigenvector centrality, similar in many
ways to the standard one, but with an important change.  We define the
centrality of node~$j$ to be the sum of the centralities of its neighbors
as before, but the neighbor centralities are now calculated \emph{in the
  absence of node~$j$}.  This is a natural definition in many ways---when I
ask my neighbors what their centralities are in order to calculate my own,
I want to know their centrality due to their other neighbors, not myself.
This modified eigenvector centrality has the desirable property that when
typical degrees are large, so that the exclusion or not of any one node
makes little difference, its value will tend to that of the standard
eigenvector centrality.  But in sparser networks of the kind that can give
problems, it will be different from the standard measure and, as we will
see, better behaved.

Our centrality measure can be calculated using the Hashimoto or
nonbacktracking matrix~\cite{Hashimoto89,Krzakala13}, which is defined as
follows.  Starting with an undirected network with $m$~edges, one first
converts it to a directed one with $2m$ edges by replacing each undirected
edge with two directed ones pointing in opposite directions.  The
nonbacktracking matrix~$\mat{B}$ is then the $2m\times2m$ non-symmetric
matrix with one row and one column for each directed edge~$i\to j$ and
elements
\begin{equation}
B_{k\to l,i\to j} = \delta_{jk}(1-\delta_{il}),
\label{eq:nbt}
\end{equation}
where $\delta_{ij}$ is the Kronecker delta.  Thus a matrix element is equal
to one if edge~$i\to j$ points into the same vertex that edge~$k\to l$
points out of and edges~$i\to j$ and $k\to l$ are not pointing in opposite
directions between the same pair of vertices, and zero otherwise.  Note
that, since the nonbacktracking matrix is not symmetric, its eigenvalues
are in general complex, but the largest eigenvalue is always real, as is
the corresponding eigenvector.

The element~$v_{i\to j}$ of the leading eigenvector of the nonbacktracking
matrix now gives us the centrality of vertex~$i$ ignoring any contribution
from~$j$, and the full nonbacktracking centrality~$x_j$ of
vertex~$j$ is defined to be the sum of these centralities over the
neighbors of~$j$:
\begin{equation}
x_j = \sum_i A_{ij} v_{i\to j}.
\label{eq:nbc}
\end{equation}
In principle one can calculate this centrality directly by calculating the
leading eigenvector of~$\mat{B}$ and then applying Eq.~\eqref{eq:nbc}.  In
practice, however, one can perform the calculation faster by making use of
the so-called Ihara (or Ihara--Bass) determinant formula, from which it can
be shown~\cite{Krzakala13} that the vector~$\vec{x}$ of centralities is
equal to the first $n$ elements of the leading eigenvector of the
$2n\times2n$ matrix
\begin{equation}
\setlength{\arraycolsep}{6pt}
\mat{M} = 
\begin{pmatrix}
\mat{A} & \mat{I}-\mat{D} \\ \mat{I} & \mat{0} \end{pmatrix},
\end{equation}
where $\mat{A}$ is the adjacency matrix as previously, $\mat{I}$ is the
$n\times n$ identity matrix, and $\mat{D}$ is the diagonal matrix with the
degrees of the vertices along the diagonal.  Since $\mat{M}$ only has
marginally more nonzero elements than the adjacency matrix itself ($2m+2n$
for a network with $m$ edges and $n$ vertices, versus $2m$ for the
adjacency matrix), finding its leading eigenvector takes only slightly
longer than the calculation of the ordinary eigenvector centrality.

To see that the nonbacktracking centrality can indeed eliminate the
localization transition, consider again our random-graph-plus-hub model
and, as before, let us first consider the random graph on its own, without
the hub.  Our goal will be to calculate the leading eigenvalue of the
nonbacktracking matrix for this random graph and then demonstrate that no
other eigenvalue ever surpasses it even when the hub is added into the
picture, and hence that there is no transition of the kind that occurs with
the standard eigenvector centrality.

Since all elements of the nonbacktracking matrix are real and nonnegative,
the leading eigenvalue and eigenvector satisfy the Perron--Frobenius
theorem, meaning the eigenvalue is itself real and nonnegative as are all
elements of the eigenvector for appropriate choice of normalization.  Note
moreover that at least one element of the eigenvector must be nonzero, so
the average of the elements is strictly positive.

Making use of the definition of the nonbacktracking matrix in
Eq.~\eqref{eq:nbt}, the eigenvector equation $z\vec{v} = \mat{B}\vec{v}$
takes the form
\begin{align}
z v_{k\to l} &= \sum_{i\to j} B_{k\to l,i\to j} v_{i\to j}
  = \sum_{i\to j} \delta_{jk} (1-\delta_{il}) v_{i\to j} \nonumber\\
 &= \sum_{ij} A_{ij} \delta_{jk} (1-\delta_{il}) v_{i\to j}
  = \sum_i A_{ik} (1-\delta_{il}) v_{i\to k}
\end{align}
or
\begin{equation}
z v_{j\to l} = \sum_{i(\ne l)} A_{ij} v_{i\to j},
\label{eq:eigen}
\end{equation}
where we have changed variables from~$k$ to $j$ for future convenience.
Expressed in words, this equation says that $z$~times the centrality of an
edge emerging from vertex~$j$ is equal to the sum of the centralities of
the other edges feeding into~$j$.  For an uncorrelated, locally tree-like
random graph of the kind we are considering here, i.e.,~a network where the
source and target of a directed edge are chosen independently and there is
a vanishing density of short loops, the centralities on the incoming edges
are drawn at random from the distribution over all edges---the fact that
they all point to vertex~$j$ has no influence on their values in the limit
of large graph size.  Bearing this in mind, let us calculate the
average~$\av{v}$ of the centralities~$v_{j\to l}$ over all edges in the
network, which we do in two stages.  First, making use of
Eq.~\eqref{eq:eigen}, we calculate the sum over all edges originating at
vertices~$j$ whose degree~$k_j$ takes a particular value~$k$:
\begin{align}
z &\sum_{\substack{j\to l:\\ k_j=k}} v_{j\to l}
  = z \!\sum_{jl:k_j=k}\! A_{jl} v_{j\to l}
  = \sum_{jl:k_j=k}\! A_{jl} \sum_{i(\ne l)} A_{ij} v_{i\to j} \nonumber\\
 &= \sum_{ij:k_j=k}\! A_{ij} v_{i\to j} \sum_{l(\ne i)} A_{jl}
  = (k-1) \!\sum_{ij:k_j=k}\! A_{ij} v_{i\to j} \nonumber\\
 &= \av{v} (k-1) \!\sum_{ij:k_j=k}\! A_{ij}
  = \av{v} (k-1) k n_k,
\end{align}
where $n_k$ is the number of vertices with degree~$k$ and we have in the
third line made use of the fact that $v_{i\to j}$ has the same distribution
as values in the graph as whole to make the replacement $v_{i\to
  j}\to\av{v}$ in the limit of large graph size.

Now we sum this expression over all values of~$k$ and divide by the total
number of edges~$2m$ to get the value of the average vector
element~$\av{v}$:
\begin{equation}
z \av{v} = {\av{v}\over2m} \sum_{k=0}^\infty (k-1) k n_k
  = \av{v} {\av{k^2}-\av{k}\over\av{k}}.
\label{eq:average}
\end{equation}
Thus for any vector~$\vec{v}$ we must either have $\av{v}=0$, which as we
have said cannot happen for the leading eigenvector, or
\begin{equation}
z = {\av{k^2}-\av{k}\over\av{k}}.
\end{equation}
For the particular case of the Poisson random graph under consideration
here, this gives a leading eigenvalue of $z=c$, the average degree.

This result has been derived previously by other means~\cite{Krzakala13}
but the derivation given here has the advantage that it is easy to adapt to
the case where we add a hub vertex to the network.  Doing so adds just a
single term to Eq.~\eqref{eq:average} thus:
\begin{equation}
z\av{v} = {\av{v}\over2m} \Biggl[ \sum_{k=0}^\infty (k-1) k n_k
          + (d-1)d \Biggr],
\end{equation}
where $d$ is the degree of the hub, as previously.  Hence the leading
eigenvalue is
\begin{equation}
z = {(n-1) \bigl( \av{k^2}-\av{k} \bigr) + (d-1)d\over2m}.
\end{equation}
For constant~$d$ and constant (or growing) average degree, however, the
term in~$d$ becomes negligible in the limit of large~$n$ and we recover the
same result as before $z=c$.

Thus no new leading eigenvalue is introduced by the hub in the case of the
nonbacktracking matrix, and there is no phase transition as eigenvalues
cross for any value of~$d$.

It is worth noting, however, that there are other mechanisms by which
high-lying eigenvalues can be generated.  For instance, if a network
contains a large clique (a complete subgraph in which every node is
connected to every other) it can generate an outlying eigenvalue of
arbitrary size, as we can see by making use of the so-called
Collatz--Wielandt formula, a corollary of the Perron--Frobenius theorem
that says that for any vector~$\vec{v}$ the leading eigenvalue satisfies
\begin{equation}
z \ge \min_{i : v_i \neq 0} \frac{[\mat{B}\vec{v}]_i}{v_i}.
\end{equation}
Choosing a $\vec{v}$ whose elements are one for edges within the clique and
zero elsewhere, we find that a clique of size~$k$ implies $z \ge k-2$,
which can supersede any other leading eigenvalue for sufficiently
large~$k$.  The corresponding eigenvector is localized on the clique
vertices, potentially causing trouble once again for the eigenvector
centrality.  This localization on cliques would be an interesting topic for
further investigation.

\subsection{Numerical results}
As a test of our nonbacktracking centrality, we show in the lower two
panels of Fig.~\ref{fig:hub} results for the same networks as the top two
panels.  As the figure makes clear, the measure now remains well behaved in
the regime beyond the former position of the localization
transition---there is no longer a large jump in the value of the centrality
on the hub or its neighbors as we pass the transition.  Similarly, the
dashed curves in Fig.~\ref{fig:transition} show the inverse participation
ratio for the nonbacktracking centrality and again all evidence of
localization has vanished.

\begin{figure*}
\begin{center}
\subfigure[\ Eigenvector centrality]{\includegraphics[width=8cm]{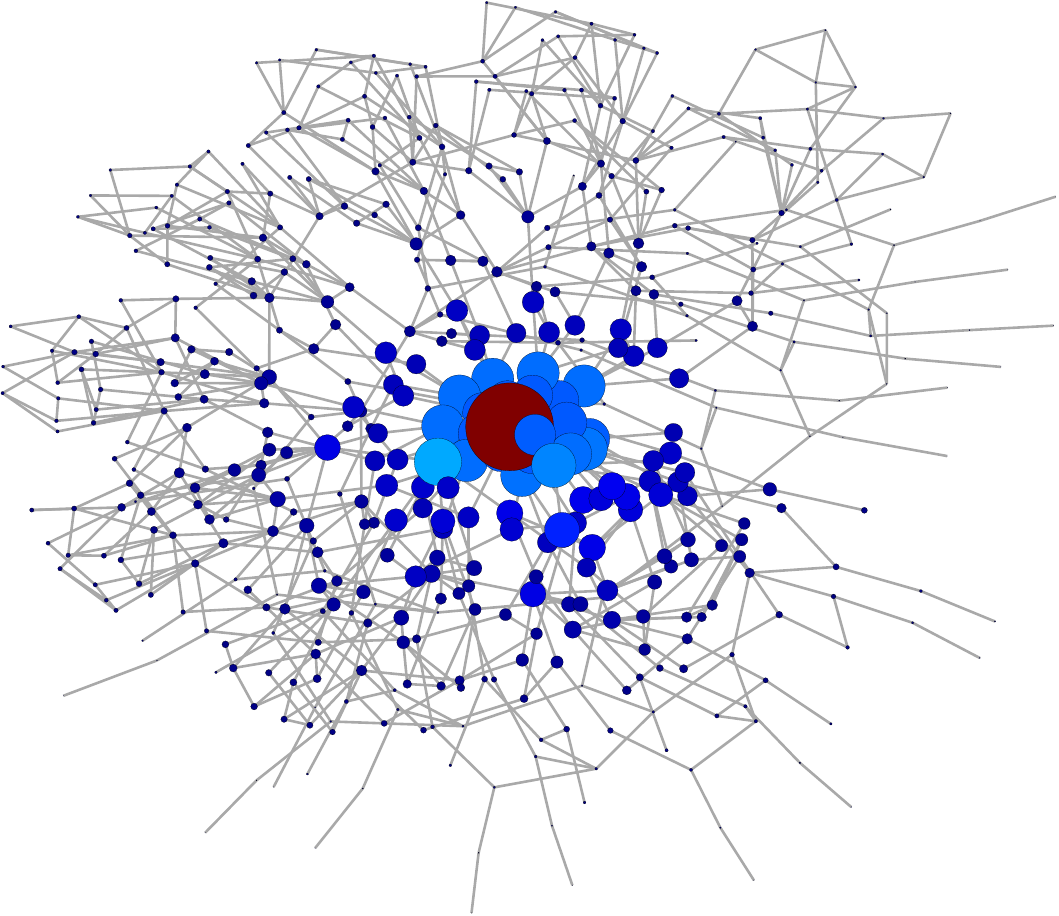} }
\subfigure[\ Nonbacktracking centrality]{\includegraphics[width=8cm]{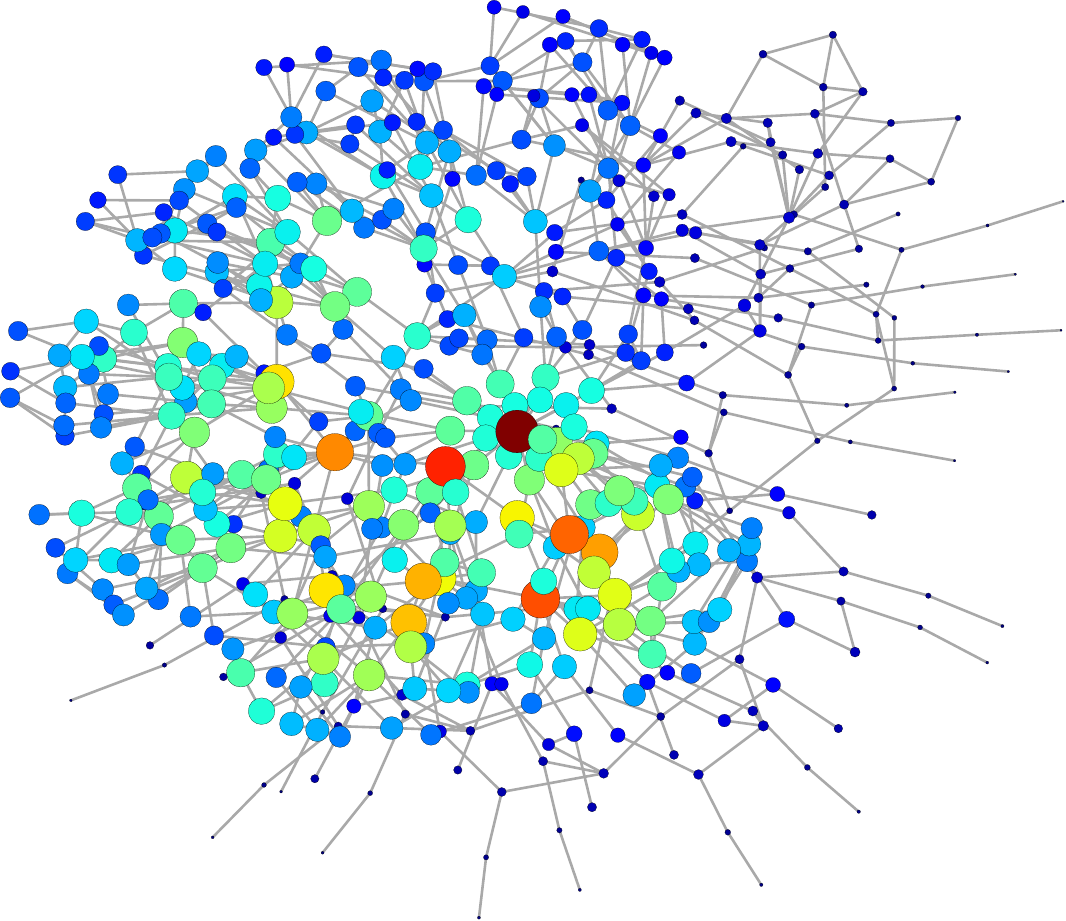}}
\end{center}
\caption{(Color online) Eigenvector and nonbacktracking centralities for
  the electronic circuit network from Table~\ref{tab:power}.  Node sizes
  are proportional to centrality (and color also varies with centrality).}
\label{fig:circuit}
\end{figure*}

\begin{table}
\centering
\setlength{\tabcolsep}{3pt}
\begin{tabular}{llrcc}
&         &                           &             & Non- \\
& Network & \multicolumn{1}{c}{Nodes} & Eigenvector & backtracking \\ \hline
\begin{rotate}{90}
\hbox{\hspace{-3.6em}Synthetic}
\end{rotate}
& Planted hub, $d=70$ & 1\,000\,001 & $2.6 \times 10^{-6} $& $1.4 \times 10^{-6}$\\
& Planted hub, $d=120$ & 1\,000\,001 & 0.2567 & $1.4 \times 10^{-6}$ \\
& Power law, $\alpha = 2.1$ & 1\,000\,000 & 0.0089 & 0.0040 \\
& Power law, $\alpha = 2.9$ & 1\,000\,000 & 0.2548 & 0.0011 \\ \hline
\begin{rotate}{90}
\hbox{\hspace{-5.2em}Empirical}
\end{rotate}
& Physics collaboration & 12\,008 & 0.0039 & 0.0039 \\
& Word associations & 13\,356 & 0.0305 & 0.0075 \\
& Youtube friendships & 1\,138\,499 & 0.0479 & 0.0047 \\
& Company ownership & 7\,253 & 0.2504 & 0.0161 \\
& Ph.D. advising & 1\,882 & 0.2511 & 0.0386 \\
& Electronic circuit & 512 & 0.1792 & 0.0056 \\ 
& Amazon & 334\,863 & 0.0510 & 0.0339
\end{tabular}
\caption{Inverse participation ratio for a variety of networks calculated
  for traditional eigenvector centrality and the nonbacktracking version.
  The first four networks are computer-generated, as described in the text.
  The remainder are, in order: a network of coauthorships of papers in
  high-energy physics~\cite{LKF05}, word associations from the Free
  Online Dictionary of Computing~\cite{BMZ02}, friendships between
  users of the Youtube online video service~\cite{Mislove07}, a network of
  which companies own which others~\cite{Norlen02}, academic advisors and
  advisees in computer science~\cite{DMB11}, electronic circuit 838 from
  the ISCAS 89 benchmark set~\cite{Milo04b}, and a product co-purchasing
  network from the online retailer Amazon.com~\cite{LKF05}.}
\label{tab:power}
\end{table}

The inverse participation ratio also provides a convenient way to test for
localization in other networks, both synthetic and real.
Table~\ref{tab:power} summarizes results for eleven networks, for both the
traditional eigenvector centrality and the nonbacktracking version.  The
synthetic networks are generated using the random-graph-plus-hub model of
this paper and the configuration model with power-law degree distribution,
and in each case there is evidence of localization in the eigenvector
centrality in the regimes where it is expected and not otherwise, but no
localization at all, in any case, for the nonbacktracking centrality.  A
similar picture is seen in the real-world networks---typically either
localization in the eigenvector centrality but not the nonbacktracking
version, or localization in neither case.  Figure~\ref{fig:circuit}
illustrates the situation for one of the smaller real-world networks, where
the values on the highest-degree vertex and its neighbors are
overwhelmingly large for the eigenvector centrality (left panel) but not
for the nonbacktracking centrality (right panel).

\section{Conclusions}
In this paper we have shown that the widely used network measure known as
eigenvector centrality fails under commonly occurring conditions because of
a localization transition in which most of the weight of the centrality
concentrates on a small number of vertices.  The phenomenon is particularly
visible in networks with high-degree hubs or power-law degree
distributions, which includes many important real-world examples.  We
propose a new spectral centrality measure based on the nonbacktracking
matrix which rectifies the problem, giving values similar to the standard
eigenvector centrality in cases where the latter is well behaved, but
avoiding localization in cases where the standard measure fails.  The new
measure is found to give significant decreases in localization on both
synthetic and real-world networks.  Moreover, the new measure can be
calculated almost as quickly as the standard one, and hence is practical
for the analysis of very large networks of the kind common in recent
studies.

The nonbacktracking centrality is not the only possible solution to the
problem of localization.  For example, in studies of other forms of
localization in networks it has been found effective to introduce a
regularizing ``teleportation'' term into the adjacency and similar
matrices, i.e.,~to add a small amount to every matrix element as if there
were a weak edge between every pair of vertices~\cite{ACBL13,QR13}.  This
strategy is reminiscent of Google's PageRank centrality
measure~\cite{BP98}, a popular variant of eigenvector centrality that
includes such a teleportation term, and recent empirical studies suggest
that PageRank may be relatively immune to localization~\cite{Ermann13}.  It
would be a worthwhile topic for future research to develop theory similar
to that presented here to describe localization (or lack of it) in PageRank
and related measures.

\begin{acknowledgments}
  The authors thank Cris Moore, Elchanan Mossel, Raj Rao Nadakuditi,
  Romualdo Pastor-Satorras, Lenka Zdeborov\'a, and Pan Zhang for useful
  conversations.  This work was funded in part by the National Science
  Foundation under grants DMS--1107796 and DMS--1407207 and by the Air
  Force Office of Scientific Research (AFOSR) and the Defense Advanced
  Research Projects Agency (DARPA) under grant FA9550--12--1--0432.
\end{acknowledgments}

\bibliographystyle{apsrev}

\end{document}